\documentclass[twocolumn, doublespace, 12pt fonts]{aastex6}
\usepackage{amsmath}

\shorttitle{GRB 170714A}
\shortauthors{Hou et al.}

\begin{document}

\title{X-ray light curve in GRB 170714A: evidence for quark star?}
\author{Shu-Jin Hou\altaffilmark{1, 2, 3}, Tong Liu\altaffilmark{4}, Ren-Xin Xu\altaffilmark{5}, Hui-Jun Mu\altaffilmark{4}, Cui-Ying Song\altaffilmark{4}, Da-Bin Lin\altaffilmark{3}, and Wei-Min Gu\altaffilmark{4}}

\altaffiltext{1}{College of Physics and Electronic Engineering, Nanyang Normal University, Nanyang, Henan 473061, China}
\altaffiltext{2}{Key Laboratory for the Structure and Evolution of Celestial Objects, Chinese Academy of Sciences, Kunming, Yunnan 650011, China}
\altaffiltext{3}{Laboratory for Relativistic Astrophysics, Department of Physics, Guangxi University, Nanning, Guangxi 530004, China}
\altaffiltext{4}{Department of Astronomy, Xiamen University, Xiamen, Fujian 361005, China; tongliu@xmu.edu.cn}
\altaffiltext{5}{School of Physics, Peking University, Beijing 100871, China; r.x.xu@pku.edu.cn}

\email{tongliu@xmu.edu.cn}

\begin{abstract}
Two plateaus and one following bump in the X-ray light curve of GRB 170714A have been detected by the \textit{Swift}/X-Ray Telescope, which could be very meaningful for the central engine of gamma-ray bursts (GRBs), implying that the origin of this burst might be different from that of other ultra-long GRBs. We propose that merging two neutron stars into a hyper-massive quark star (QS) and then collapsing into a black hole (BH), with a delay time around $10^4$~s, could be responsible for those X-ray components. The hyper-massive QS is initially in a fluid state, being turbulent and differentially rotating, but would be solidified and release its latent heat injected into the GRB fireball (lasting about $10^3$~s during the liquid-solid phase transition). Magnetic field as high as $\sim 10^{15}$~G could be created by dynamo action of the newborn liquid QS, and a magnetar-like central engine (after solidification) supplies significant energy for the second plateau. More energy could be released during a fall-back accretion after the post-merger QS collapses to a BH, and the X-ray bump forms. This post-merger QS model might be tested by future observations, with either advanced gravitational wave detectors (e.g., advanced LIGO and VIRGO) or X-ray/optical telescopes.
\end{abstract}


\section{Introduction\label{sec:intro}}

It remains a mystery about many aspects of the gamma-ray bursts (GRBs), including the central engine and the radiation mechanism (for reviews, see e.g., Zhang 2011; Kumar \& Zhang 2015). According to the duration of the prompt emission, GRBs are classified into two categories: long- and short-duration GRBs (LGRBs and SGRBs). They are generally related to the collapsars (see reviews by Woosley \& Bloom 2006) and the compact binary [neutron star (NS)-NS or NS-black hole (BH)] mergers (e.g., Eichler et al. 1989; Paczy\'nski 1991; Narayan et al. 1992), respectively. However, some recent observations support the existence of new subclasses of GRBs: long-short GRBs (e.g., Gehrels et al. 2006; Zhang et al. 2007, 2009) and ultra-long GRBs (ULGRBs, see e.g., Gendre et al. 2013; Levan et al. 2014; Zhang et al. 2014; Greiner et al. 2015; Ioka et al. 2016).

The typical GRB X-ray light curve is generally divided into five distinct phases, i.e., steep decay, shallow decay, normal decay, the late steep decay, and X-ray flares (e.g., Zhang et al. 2006; Nousek et al. 2006). The internal X-ray plateaus often appear in the X-ray afterglow of LGRBs and SGRBs (see e.g., Rowlinson et al. 2010; L\"{u} \& Zhang 2014; L\"{u} et al. 2015; Gao et al. 2016b) and might be related to long-lasting activities of the central engines (e.g., Troja et al. 2007). The energy injection from magnetars or quark stars (QSs) had been proposed to explain these phenomena (e.g., Dai \& Lu 1998; Zhang \& M{\'e}sz{\'a}ros 2001; Paczy\'nski \& Haensel 2005; Fan \& Xu 2006; Staff et al. 2008; Lyons et al. 2010; Wu et al. 2014; Piro et al. 2014; Gao et al. 2016b; Li et al. 2016, 2017; Beniamini \& Mochkovitch 2017).

\begin{figure}\label{Fig_1}
\centering
\includegraphics[angle=0,scale=0.8]{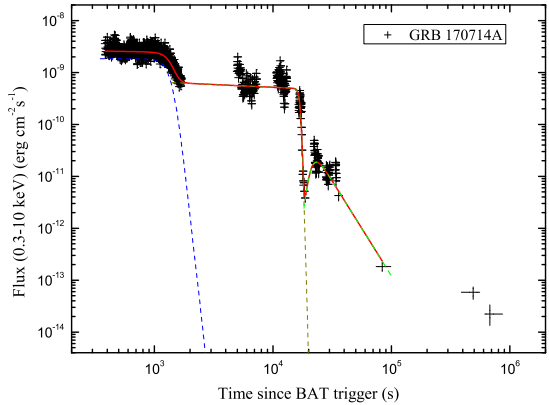}
\caption{XRT (black) light curve of GRB 170714A. Three smooth broken power-law functions are used to fit the data from $\sim$ 400 s to $\sim$ 83786 s. The solid line and the dashed line represent the total and each component fitting line. The parameters of these three parts ($\alpha$, $\beta$, $t_{b}$) are (0.02, 20.83, 1473 s), (0.11, 80.15, 17223 s), and (-23.25, 3.66, 20288 s), respectively.}
\end{figure}

It is worth noting that, after the binary NS merger, a new compact star is formed. It may not immediately collapse into a BH, depending on the equation of state (EoS) of NS matter, due to rotation (see review by Bartos et al. 2013). In fact, the EoS of dense matter at a few nuclear densities is a great challenge in physics and astronomy, and it is still a matter of debate that the fundamental degree of freedom of supranuclear matter is either hadron or quark (Weber et al. 2013). Nevertheless, it is addressed, from an astrophysical point of view, that a pulsar-like compact star could be in a solid state of quark matter (Xu 2003), with strangeon (former name: quark-cluster) as the constituent because of strong coupling so that quarks are localized there (Lai \& Xu 2017). A hot strangeon star would be in a liquid state, but could be phase-converted to a solid state when its temperature is lower than $\sim 1$ MeV. It is proposed that GRB X-ray plateau could be understood by considering the solidification of newborn strangeon stars with latent heat released as energy injection to the GRB afterglow (Dai et al. 2011). Note that the state equation of strangeon matter is so stiff that massive pulsars ($\gtrsim 2~M_\odot$) could be naturally explained (Lai \& Xu 2009, 2017). It is worth noting that the strangeon star model pass the examination of dynamical test of tidal polarizablility, while mergers of two strangeon stars and accompanying electromagnetic radiations had been visited based on the multi-band observations of GW 170817 (Lai et al. 2017).

In ULGRB observations, the various profiles of the X-ray light curves appear. For example, the X-ray light curve of GRB 101225 can be fitted by two smooth exponential functions (e.g., Campana et al. 2011), the X-ray afterglow of GRB 111209A can be represented by three-segment functions with a supernova (SN)-like bump (e.g., Gendre et al. 2013; Stratta et al. 2013; Greiner et al. 2015; Ioka et al. 2016; Gao et al. 2016a; Liu et al. 2018), GRB 121027A has a large X-ray bump superimposed on the shallow decay (e.g., Hou et al. 2014a; Zhang et al. 2014; Wu et al. 2013), and GRB 130925A has a lot of giant flares superimposed on the shallow decay (e.g., Hou et al. 2014b; Piro et al. 2014). The significant peculiarity of these bursts might be related to the certain processes or mechanisms of the GRB central engines. In various guises, they are served the single origin, i.e., BH hyperaccretion or magnetar resulted by the collapsars or compact binary mergers. Ioka et al. (2016) investigated three candidates of the ULGRB central engine, i.e., the blue supergiant collapsars, newborn magnetars, and white dwarf tidal disruption, on GRB 111209A associated with SN 2011kl. They found that all three models can explain this burst, although the SN-like bump require that the spin-down time of the magnetar is a hundred times longer than the timescale of the GRB. Liu et al. (2018) also tested the initial masses and metallicities of the progenitor stars of ULGRBs by using BH hyperaccretion inflow-outflow model. Furthermore, Gao et al. (2016a) suggested that GRB 111209A/SN 2011kl may be originated from the BH hyperaccretion process with the Blandford-Znajek (BZ) mechanism (Blandford \& Znajek 1977) and Blandford-Payne machanism (Blandford \& Payne 1982).

Recently, another ULGRB GRB 170714A was observed by the \textit{Swift} telescope. Its X-ray light curve might be composed of two plateaus and one bump. This characteristic feature challenges all the known models of the central engine. Therefore, we propose to interpret three X-ray components of GRB 170714A by using the phase transition of a QS, the QS spin-down process and fall-back accretion into a BH, respectively. In Section 2, the data analysis is shown. We describe our model in Section 3. Summary is drawn in Section 4.

\section{Data\label{sec:data}}

GRB 170714A was discovered at $T_0 = 12:25:32$ UT on 2017 July 14 by the Burst Alert Telescope (BAT) on board {\it Swift} (D'Ai et al. 2017) and accurately located by X-Ray Telescope (XRT) at a position of $\alpha$ = 02$^{\rm h}$17$^{\rm m}$23.95$^{\rm s}$, $\delta$ = -1$^{\circ}$59$'$24.4$''$ (J2000), with an uncertainty of $5.^{''}0$ (Evans et al. 2017). The redshift is $z= 0.793$ (de Ugarte Postigo et al. 2017). The mask-weighted light curve of prompt emission shows no obvious pulse and only continuous weak emission (Palmer et al. 2017). Then it is difficult to estimate $T_{90}$. The time integrated spectrum from $T_0-73$ s to $T_0+464.4$ s is best fitted by a simple power-law model and the fluence in the $15 - 150 \rm~keV$ energy band is $(2.8\pm0.3) \times 10^{-6}$ erg cm$^{-2}$ (Palmer et al. 2017), yielding an isotropic gamma-ray energy release about $(1.58 \pm 0.08) \times10^{52}$ erg.

\begin{deluxetable*}{cccccc}\label{table1}
\tablenum{1}
\tablecaption{Fitting results of X-ray light curves of GRB 170714A}
\tabletypesize{\scriptsize}
\tablewidth{0pt}
\tablehead{\colhead{} & \colhead{$\alpha$} & \colhead{$\beta$} & \colhead{$F_{\rm 0}~(\rm erg~cm^{-2}~s^{-1})$} & \colhead{$t_{\rm b}~(\rm s)$} & \colhead{$L_{\rm b}~(\rm erg~s^{-1})$}}
\startdata
1st plateau	&	0.02 	$\pm$	0.05 	&	20.83 	$\pm$	1.85 	& (1.81 	$\pm$	0.09)$\times10^{-9}$	&	1473	$\pm$	15 	& (5.47	$\pm$	0.27)$\times10^{48}$\\
2nd plateau	&	0.11 	$\pm$	0.05 	&	80.15 	$\pm$	5.16 	& (4.91 	$\pm$	0.21)$\times10^{-10}$	&	17223 	$\pm$	37	& (1.48	$\pm$	0.06)$\times10^{48}$\\
Bump	    &	-23.25 	$\pm$	3.68 	&	3.66 	$\pm$	0.22 	& (4.14 	$\pm$	0.32)$\times10^{-11}$	&	20288 	$\pm$	343 & (1.25	$\pm$	0.10)$\times10^{47}$\\
\enddata
\end{deluxetable*}

The observation of XRT on this burst began at $T_0$ + 392.7 s (D'Avanzo et al. 2017). Figure 1 shows the XRT light curve in the $0.3 - 10 \rm~keV$  band (Evans et al. 2009). Unfortunately, due to the satellite motion, there is no data in 5 time gaps, i.e., from $\sim$ 1700 s to $\sim$ 5000 s, $\sim$ 7300 s to $\sim$ 10700 s, $\sim$ 13100 s to $\sim$ 16500 s, and $\sim$ 35800 s to $\sim$ 83700 s. However, the contours of light curve can still be inferred. The XRT light curve of GRB 170714A is unusual. Two plateaus and one following bump are dominant in the X-ray light curve.

Three smooth broken power-law functions are used to fit the data from $\sim$ 400 s to $\sim$ 83786 $\rm~s$. The first plateau decays as a power law with the temporal index $\alpha_{1} \sim 0.02$ and $\beta_{1} \sim 20.83$ until the steep decay at $t_{b,1} \sim 1473$ s and lasts $\sim 1700$ s. The temporal indexes of the second plateau are $\alpha_{2} \sim 0.11$ and $\beta_{2} \sim 80.15$, respectively, and the break time is about $t_{b,2} \sim 17223$ s. This plateau is obviously superimposed with some flares. Interestingly, the end of second plateau has a deep dip, which implies that a giant bump is following. There still exists several flares. The values of $\alpha_{3}$ and $\beta_{3}$ are about -23.25 and 3.66. The peak time is $\sim 20288$ s. All the fitting results are reported in Table 1, including $\alpha$, $\beta$, the X-ray flux $F_0$, the break time $t_b$, and the isotropic luminosity $L_b$ of three components. We can roughly estimate the energy of the three components being about $3\times10^{51} \rm~erg$, $10^{52} \rm~erg$, and $10^{51} \rm~erg$, respectively.

\section{Model\label{sec:model}}

Generally, the compact binary merger can release a large amount of energy to power the prompt emission of SGRBs. After mergers, a new compact star is born, and can be either a massive magnetar, a QS or a BH. For GRB 170714A, we assume that two NSs merger occurs in the center, then a QS forms with mass about 3 $M_\odot$. The massive and highly rotating QS may be in the liquid phase owing to the severe conditions in the merger process. This phase is unstable. After an initial cooling stage due to neutrino and photon emission, the phase of matter inside the new QS will quickly change to solid state. It should be noted that this process will be accompanied by an energetic release. Same as a magnetar, the solidified QS can release its rotational energy via the strong magnetic dipole radiation. When the QS spins down and the self-gravity cannot be against, it will collapse into a BH.

According to the solid QS model, the depth of the potential $U_{0}$ usually takes 100 MeV (Dai et al. 2011) and the ratio of melting heat to potential $f$ is between 0.01 and 0.1 (Lai et al. 2009). Then the released energy per baryon during the phase transition can be estimated as
\begin{equation}
E \sim f U_{0}\approx 1 - 10 ~\rm MeV.
\end{equation}
Here the mass of a QS is assumed to be 3 $M_\odot$ ($\sim 6\times 10^{33} ~\rm g$), then the number of baryon $n$ is about $3 \times 10^{57}$.
During the phase transition process, the total energy released by the QS, $E_{1}$, is roughly estimated as
\begin{equation}
E_{1}= n E \sim n f U_{0} \approx 5 \times 10^{51} - 5 \times 10^{52} ~\rm erg.
\end{equation}
$E_1$ is in the magnitude of the energy required by the shallow decay phase of a GRB. Here we assume the blackbody radiation luminosity $L_{1}$, which can be represented as
\begin{equation}
L_{1}=\sigma T^4 4 \pi R^2,
\label{eq1}
\end{equation}
where $\sigma$ is Stefan-Boltzman constant, $R$ is the radius of the new-born QS and $T$ is the temperature of blackbody radiation. The radiation timescale $t_1$ can be express as $E_{1}/L_{1}$. If the temperature $kT$ is 1 $\rm MeV$, the timescale would be around thousands of seconds, which agrees well with the typical X-ray plateau of GRBs. More important, the luminosity $L_{1}$ can be roughly considered as a constant because the temperature almost remains unchanged. So the corresponding light curve would appear as a plateau. Of course, there must exist an efficiency from $L_{1}$ to the isotropic luminosity of the plateau. It should be emphasized that the released energy in the initial cooling stage and the phase transition process should be injected into the GRB jet and power the nonthermal radiation.

The spin-down of magnetar is widely used to explain the plateaus in both LGRBs and SGRBs. We consider that the nature of the newborn QS is similar to that of the magnetar. Then the characteristic spin-down luminosity of QS, $L_{2}$, can be expressed as (e.g., Zhang \& M\'{e}sz\'{a}ros 2001; L\"{u} \& Zhang 2014)
\begin{equation}
L_{2} \approx 10^{49}B_{p,15}^2 P_{0,-3}^{-4} R_6^6~{\rm erg~s^{-1}},
\end{equation}
and the characteristic spin-down timescale $\tau $ of the QS can be written as
\begin{equation}
\tau \approx  2 \times 10^4 I_{46} B_{p,15}^{-2} P_{0,-3}^2 R_6^{-6}~{\rm s},
\end{equation}
where $I_{46}$ is the dimensionless moment of inertia of a QS ($I \sim 10^{46} ~\rm g~cm^2$ for massive QSs, see e.g., Li et al. 2016, 2017), $B_{p,15}$ is the dimensionless magnetic field strength, $P_{0,-3}$ is the dimensionless initial period, and $R_6$ is the dimensionless QS radius. If the isotropic luminosity of a plateau is $L_{\rm b, 2}$, we can get $L_{\rm b, 2}= \xi L_2$, where $\xi$ is a coefficient by considering the radiation efficiency and the beaming factor and in 0-1 range.

A massive QS should finally collapse into a BH if the centrifugal force fails to contend with gravity. Since the ejecta emerged from the merger events falls back, a BH hyperaccretion system forms.

Wu et al. (2013) proposed that a BH fall-back accretion with the BZ mechanism powering a jet can interpret a giant X-ray bump of GRB 121027A in the collapsar scenario for LGRBs. Hou et al. (2014a) analyzed the variability of the giant X-ray bump in GRB 121027A and suggested that a jet precession in the BH hyperaccretion framework can explain the variability. Recently, similar to our motivation, Chen et al. (2017) found that a small X-ray bump follows the plateau in GRB 070110, which can be interpreted by a fall-back accretion onto BH collapsed from a spin-down magnetar. They considered that the bump can be regarded as an evidence of the magnetar powering the internal plateau.

We consider that the BZ mechanism may dominate in this accretion process. The BZ luminosity $L_{\rm BZ}$ can be written as (e.g., Lee et al. 2000a,b; Liu et al. 2018)
\begin{equation}
L_{\rm BZ} =f(a_*)c R_{\rm g}^2 \frac{B_{\rm in}^2}{8 \pi},
\end{equation}
where $a_*$ is the dimensionless BH spin parameter, $f (a_*)$ is a factor depending on the specific configuration of the magnetic field, $R_{\rm g}=GM_{\rm BH}/c^{2}$ is the Schwarzschild radius, $M_{\rm BH}$ is the BH mass, and $B_{\rm in}$ is the poloidal magnetic field strength near the BH horizon. If the isotropic luminosity of the bump is $L_{\rm b, 3}$, then $L_{\rm b,3}= \zeta L_{\rm BZ}$. The coefficient $\zeta$ also includes the radiation efficiency and the beaming factor and its range is 0-1.

Furthermore, according to the balance between ram pressure of the innermost part of the disk $P_{\rm in}$ and the magnetic pressure on the BH horizon (e.g., Liu et al. 2017a,c), one has
\begin{equation}
\frac{B_{\rm in}^{2}}{8\pi}=P_{\rm in} \sim \rho_{\rm in} c^{2}\sim\frac{\dot{M}_{\rm in}c}{4\pi R_{\rm H}^{2}},
\end{equation}
where $R_{\rm H}=(1+\sqrt{1-a_*^{2}})R_{\rm g}$ is the radius of the BH horizon, and $\dot{M}_{\rm in}$ and $\rho_{\rm in}$ are the net accretion rate and density at the inner boundary of the disk, respectively. We can estimate the accreted mass, i.e., the lower limit of the mass of the ejecta from mergers, based on the above equation.

We consider that the gamma- and X-ray features of GRB 170714A can be well explained in the following scenario.

\textbf{(a) \emph{Prompt emission}} ~~In the beginning, after the merger event of two NSs, the new-born QS should undergo the initial cooling stage. If we reasonably assume that the initial temperature ($kT$) is about 30-50 MeV, and only 10$\%$ energy has been injected into the fireball then released by the photons, the gamma-ray energy of GRB 170714A is satisfied with the cooling mechanism. The smooth cooling process just corresponds to the continuous weak gamma-ray emission.

\textbf{(b) \emph{First plateau}} ~~The phase transition provides the energy to interpret the first X-ray plateau from $\sim$ 400 s to $\sim$ 1700 s. At this stage, the energy conversion efficiency is set as 0.1, then the energy of the first plateau is within the range of energy released by the phase transition. We assume that $t_1$ roughly equals to $t_{b, 1}/(1+z)$ and $R$ is takes 10 km, then the blackbody temperature can be estimated as $kT \sim 4 ~\rm MeV$, which is within the reasonable limits (e.g., Yuan et al. 2017).

\textbf{(c) \emph{Second plateau}} ~~For the plateau from $\sim$ 1700 s to $\sim$ 13000 s, when the coefficient $\xi$ takes 0.1 and the characteristic spin-down time scale $\tau \sim t_{b ,2}/(1+z)$, the initial period $P$ and the magnetic field strength $B$ can be inferred to be $\sim$ 1.2 ms and $\sim 1.6 \times 10^{15}$ $\rm G$ by Equations (4) and (5), which are all in the reasonable value range of massive newborn QSs (e.g., Li et al. 2016).

\textbf{(d) \emph{Bump}} ~~For the bump from $\sim$ 19000 s to $\sim$ 80000 s, we can estimate $B_{\rm in}$ by Equation (6). If $\zeta$, $a_*$, and $f (a_*)$ take 0.1, 0.9, and 1 (e.g., Liu et al. 2015), and corresponding the observations, $L_{\rm BZ}$ is about $1.25 \times10^{48}$ erg $\rm s^{-1}$, then we get $B_{\rm in}$ is about $3\times10^{13} \rm~ G$, which is suitable with the QS spin-down process. Form Equation (7), we derive the accretion mass of $\sim 0.15 ~ M_{\odot}$, which is lower than the ejecta mass resulted by the NS-NS merger simulations (e.g., Dietrich et al. 2015).

\section{Summary}

In this paper, we studied the X-ray features of GRB 170714A and discussed their possible origins. There are two plateaus and one bump superimposed on the X-ray afterglow, which is quite different from the normal X-ray afterglow. We proposed that the fast cooling stage of a newborn QS after merger corresponds to the prompt emission, and the phase transition of a QS, spin-down of a QS, and BH fall-back hyperaccretion can be used to explain the three X-ray components in turn. We tested that above theoretical framework is reasonable and self-consistent. Then we considered that the X-ray multi-plateaus phase of GRBs might be the evidence of the existence of QSs.

As well as a magnetar, if a QS exists in the center of a GRB, it will collapse into a BH or keep a stable magnetar after spin-down (e.g., Bartos et al. 2013; L{\"u} \& Zhang 2014; L\"{u} et al. 2015; Chen et al. 2017). For GRB 170714A, the massive QS, $\sim 3~M_\odot$, may indeed collapse into a BH, then the BH hyperaccretion process powers an X-ray bump.

In addition, it is generally believed that the mass of the ejecta from NS-NS mergers is larger than that from QS-QS mergers, so considering that the parts of ejecta are required in the BH fall-back accretion process to effectively reignite the central engine and provide the energy of the bump, we believe that the NS-NS merger might be the progenitor of GRB 170714A. Besides, other parts of ejecta may power other potential electromagnetic counterparts like kilonovae (or mergernovae, see e.g., Li \& Paczy{\'n}ski 1998; Metzger et al. 2010; Yu et al. 2013; Metzger 2017; Song et al. 2017), so the NS-NS merger could be favorable for GRB 170714A.

In consideration of the requirements of the phase transition and the ejecta from mergers, the total mass of the progenitor of GRB 170714A may be greater than 3 $M_\odot$, which implies that the similar events are definitely rare. Even though, more samples of X-ray light curves of GRBs like GRB 170714A are still expected to reveal the secret of this issues, especially with synergy observations of gravitational wave (GW) detectors, to constrain the EoS of NSs or QSs.

Liu et al. (2017b) compared the GWs from the BH hyperaccretion processes and millisecond magnetar models as the candidates of the GRB central engines. If combining the multi-band electromagnetic signals, one might take the mystery out of the hidden GRB central engine. For the possible high-energy explosions like GRB 170714A, we can expect that the binary compact stars merging and QSs collapsing will produce GWs and X-ray multi-plateaus, which will be observed by either advanced GW detectors (e.g., advanced LIGO and VIRGO) or X-ray/optical telescopes, and our model will be further verified.

\acknowledgments

We thank Ang Li and Bing Zhang for helpful discussion. We acknowledge the use of the public data from the \textit{Swift} data archives. This work was supported by the National Basic Research Program of China (973 Program) under grant 2014CB845800, the National Natural Science Foundation of China under grants 11503011, 11473022, 11573023, 11673002, 11773007, and U1531243, the CAS Open Research Program of Key Laboratory for the Structure and Evolution of Celestial Objects (grant OP201403), the Key Scientific Research Project in Universities of Henan Province (grant 15A160001), and the Doctoral Foundation of Nanyang Normal University.


\begin{thebibliography}{99}
\bibitem[Bartos et al.(2013)]{2013CQGra..30l3001B} Bartos, I., Brady, P., \& M{\'a}rka, S.\ 2013, Classical and Quantum Gravity, 30, 123001
\bibitem[Beniamini \& Mochkovitch(2017)]{} Beniamini, P., \& Mochkovitch, R. 2017, \aap, 605A, 60
\bibitem[Blandford \& Payne(1982)]{1982MNRAS.199..883B}Blandford, R. D., \& Payne, D. G. 1982, MNRAS, 199, 883
\bibitem[Blandford \& Znajek(1977)]{1977MNRAS.179..433B} Blandford, R.~D., \& Znajek, R.~L.\ 1977, \mnras, 179, 433
\bibitem[Campana et al.(2011)]{2011Natur.480...69C} Campana, S., Lodato, G., D'Avanzo, P., et al.\ 2011, \nat, 480, 69
\bibitem[Chen et al.(2017)]{} Chen, W., Xie, W., Lei, W.-H., et al.\ 2017, \apj, 849, 119
\bibitem[D'Ai et al.(2017)]{2017GCN.21340....1D} D'Ai, A., Burrows, D.~N., Cholden-Brown, A., et al.\ 2017, GRB Coordinates Network, Circular Service, 21340, 1
\bibitem[D'Avanzo et al.(2017)]{2017GCN.21343....1D} D'Avanzo, P., D'Elia, V., Cholden-Brown, A., et al.\ 2017, GRB Coordinates Network, Circular Service, 21343, 1
\bibitem[Dai et al.(2011)]{2011SCPMA..54.1541D} Dai, S., Li, L. Y., \& Xu, R. X.\ 2011, Science China Physics, Mechanics, and Astronomy, 54, 1541
\bibitem[Dai \& Lu(1998)]{1998A&A...333L..87D} Dai, Z.~G., \& Lu, T.\ 1998, \aap, 333, L87
\bibitem[de Ugarte Postigo et al.(2017)]{2017GCN.21359....1D} de Ugarte Postigo, A., Kann, D.~A., Izzo, L., et al.\ 2017, GRB Coordinates Network, Circular Service, 21359, 1
\bibitem[Dietrich et al.(2015)]{2015PhRvD..91l4041D} Dietrich, T., Bernuzzi, S., Ujevic, M., \& Br{\"u}gmann, B.\ 2015, \prd, 91, 124041
\bibitem[Eichler et al.(1989)]{1989Natur.340..126E} Eichler, D., Livio, M., Piran, T., \& Schramm, D.~N.\ 1989, \nat, 340, 126
\bibitem[Evans et al.(2009)]{2009MNRAS.397.1177E} Evans, P.~A., Beardmore, A.~P., Page, K.~L., et al.\ 2009, \mnras, 397, 1177
\bibitem[Evans et al.(2017)]{2017GCN.21341....1E} Evans, P.~A., Goad, M.~R., Osborne, J.~P., \& Beardmore, A.~P.\ 2017, GRB Coordinates Network, Circular Service, 21341, 1
\bibitem[Fan \& Xu(2006)]{2006MNRAS.372L..19F} Fan, Y.-Z., \& Xu, D.\ 2006, \mnras, 372, L19
\bibitem[Gao et al.(2016)]{2016ApJ...826..141G} Gao, H., Lei, W.-H., You, Z.-Q., \& Xie, W.\ 2016a, \apj, 826, 141
\bibitem[Gao et al.(2016)]{2016PhRvD..93d4065G} Gao, H., Zhang, B., \& L{\"u}, H.-J.\ 2016b, \prd, 93, 044065
\bibitem[Gehrels et al.(2006)]{2006Natur.444.1044G} Gehrels, N., Norris, J.~P., Barthelmy, S.~D., et al.\ 2006, \nat, 444, 1044
\bibitem[Gendre et al.(2013)]{2013ApJ...766...30G} Gendre, B., Stratta, G., Atteia, J.~L., et al.\ 2013, \apj, 766, 30
\bibitem[Greiner et al.(2015)]{2015Natur.523..189G} Greiner, J., Mazzali, P.~A., Kann, D.~A., et al.\ 2015, \nat, 523, 189
\bibitem[Hou et al.(2014)]{2014MNRAS.441.2375H} Hou, S.-J., Gao, H., Liu, T., et al.\ 2014a, \mnras, 441, 2375
\bibitem[Hou et al.(2014)]{2014ApJ...781L..19H} Hou, S.-J., Liu, T., Gu, W.-M., et al.\ 2014b, \apjl, 781, L19
\bibitem[Ioka et al.(2016)]{2016ApJ...833..110I} Ioka, K., Hotokezaka, K., \& Piran, T.\ 2016, \apj, 833, 110
\bibitem[Kumar \& Zhang(2015)]{} Kumar, P., \& Zhang, B. 2015, Phys. Rep., 561, 1
\bibitem[Lai \& Xu(2009)]{2009MNRAS.398L..31L} Lai, X.~Y., \& Xu, R.~X.\ 2009, \mnras, 398, L31
\bibitem[Xiaoyu \& Renxin(2017)]{2017JPhCS.861a2027X} Lai, X. Y. \& Xu, R. X.\ 2017, Journal of Physics Conference Series, 861, 012027
\bibitem[Lai et al.(2017)]{2017arXiv171004964L} Lai, X.~Y., Yu, Y.~W., Zhou, E.~P., Li, Y.~Y., \& Xu, R.~X.\ 2017, Research in Astronomy and Astrophysics, in press, arXiv:1710.04964
\bibitem[Lee et al.(2000a)]{Lee2000a} Lee, H.~K., Brown, G.~E., \& Wijers, R.~A.~M.~J.\ 2000a, \apj, 536, 416
\bibitem[Lee et al.(2000b)]{Lee2000b} Lee, H.~K., Wijers, R.~A.~M.~J., \& Brown, G.~E.\ 2000b, Physics Reports, 325, 83
\bibitem[Levan et al.(2014)]{2014ApJ...781...13L} Levan, A.~J., Tanvir, N.~R., Starling, R.~L.~C., et al.\ 2014, \apj, 781, 13
\bibitem[Li \& Paczy{\'n}ski(1998)]{Li1998} Li, L.-X., \& Paczy{\'n}ski, B.\ 1998, \apjl, 507, L59
\bibitem[Li et al.(2016)]{2016PhRvD..94h3010L} Li, A., Zhang, B., Zhang, N.-B., et al.\ 2016, \prd, 94, 083010
\bibitem[Li et al.(2017)]{2017ApJ...844...41L} Li, A., Zhu, Z.-Y., \& Zhou, X.\ 2017, \apj, 844, 41
\bibitem[Liu et al.(2017)]{2017arXiv170505516L} Liu, T., Gu, W.-M., \& Zhang, B.\ 2017a, New Astronomy Reviews, 79, 1
\bibitem[Liu et al.(2015)]{2015ApJS..218...12L} Liu, T., Hou, S.-J., Xue, L., \& Gu, W.-M.\ 2015, \apjs, 218, 12
\bibitem[Liu et al.(2017)]{2017arXiv170909810L} Liu, T., Lin, C.-Y., Song, C.-Y., \& Li, A.\ 2017b, \apj, 850, 30
\bibitem[Liu et al.(2017)]{2017arXiv171000141L} Liu, T., Song, C.-Y., Zhang, B., Gu, W.-M., \& Heger, A.\ 2018, \apj, 852, 20
\bibitem[L{\"u} \& Zhang(2014)]{2014ApJ...785...74L} L{\"u}, H.-J., \& Zhang, B.\ 2014, \apj, 785, 74
\bibitem[L{\"u} et al.(2015)]{2015ApJ...805...89L} L{\"u}, H.-J., Zhang, B., Lei, W.-H., Li, Y., \& Lasky, P.~D.\ 2015, \apj, 805, 89
\bibitem[Lyons et al.(2010)]{2010MNRAS.402..705L} Lyons, N., O'Brien, P.~T., Zhang, B., et al.\ 2010, \mnras, 402, 705
\bibitem[Metzger(2017)]{Metzger2017} Metzger, B.~D.\ 2017, Living Reviews in Relativity, 20, 3
\bibitem[Metzger et al.(2010)]{Metzger2010} Metzger, B.~D., Mart{\'{\i}}nez-Pinedo, G., Darbha, S., et al.\ 2010, \mnras, 406, 2650
\bibitem[Narayan et al.(1992)]{1992ApJ...395L..83N} Narayan, R., Paczynski, B., \& Piran, T.\ 1992, \apjl, 395, L83
\bibitem[Nousek et al.(2006)]{2006ApJ...642..389N} Nousek, J.~A., Kouveliotou, C., Grupe, D., et al.\ 2006, \apj, 642, 389
\bibitem[Paczynski(1991)]{1991AcA....41..257P} Paczy\'nski, B.\ 1991, \actaa, 41, 257
\bibitem[Paczy{\'n}ski \& Haensel(2005)]{2005MNRAS.362L...4P} Paczy{\'n}ski, B., \& Haensel, P.\ 2005, \mnras, 362, L4
\bibitem[Palmer et al.(2017)]{2017GCN.21347....1P} Palmer, D.~M., Barthelmy, S.~D., Cummings, J.~R., et al.\ 2017, GRB Coordinates Network, Circular Service, 21347, 1
\bibitem[Piro et al.(2014)]{2014ApJ...790L..15P} Piro, L., Troja, E., Gendre, B., et al.\ 2014, \apjl, 790, L15
\bibitem[Rowlinson et al.(2010)]{2010MNRAS.409..531R} Rowlinson, A., O'Brien, P.~T., Tanvir, N.~R., et al.\ 2010, \mnras, 409, 531
\bibitem[Song et al.(2017)]{2017arXiv171000142S} Song, C.-Y., Liu, T., \& Li, A.\ 2017, arXiv:1710.00142
\bibitem[Staff et al.(2008)]{2008MNRAS.391..178S} Staff, J., Niebergal, B., \& Ouyed, R.\ 2008, \mnras, 391, 178
\bibitem[Stratta et al.(2013)]{2013ApJ...779...66S} Stratta, G., Gendre, B., Atteia, J.~L., et al.\ 2013, \apj, 779, 66
\bibitem[Troja et al.(2007)]{2007ApJ...665..599T} Troja, E., Cusumano, G., O'Brien, P.~T., et al.\ 2007, \apj, 665, 599
\bibitem[Weber et al.(2013)]{2013IAUS..291...61W} Weber, F., Orsaria, M., Rodrigues, H., \& Yang, S.-H.\ 2013, Neutron Stars and Pulsars: Challenges and Opportunities after 80 years, 291, 61
\bibitem[Woosley \& Bloom(2006)]{2006ARA&A..44..507W} Woosley, S.~E., \& Bloom, J.~S.\ 2006, \araa, 44, 507
\bibitem[Wu et al.(2014)]{2014ApJ...781L..10W} Wu, X.-F., Gao, H., Ding, X., et al.\ 2014, \apjl, 781, L10
\bibitem[Wu et al.(2013)]{2013ApJ...767L..36W} Wu, X.-F., Hou, S.-J., \& Lei, W.-H.\ 2013, \apjl, 767, L36
\bibitem[Xu(2003)]{2003ApJ...596L..59X} Xu, R.~X.\ 2003, \apjl, 596, L59
\bibitem[Yu et al.(2013)]{Yu2013} Yu, Y.-W., Zhang, B., \& Gao, H.\ 2013, \apjl, 776, L40
\bibitem[Yuan et al.(2017)]{2017RAA....17...92Y} Yuan, M., Lu, J.-G., Yang, Z.-L., Lai, X.-Y., \& Xu, R.-X.\ 2017, Research in Astronomy and Astrophysics, 17, 092
\bibitem[Zhang(2011)]{2011CRPhy..12..206Z} Zhang, B.\ 2011, Comptes Rendus Physique, 12, 206
\bibitem[Zhang et al.(2006)]{2006ApJ...642..354Z} Zhang, B., Fan, Y.~Z., Dyks, J., et al.\ 2006, \apj, 642, 354
\bibitem[Zhang \& M{\'e}sz{\'a}ros(2001)]{2001ApJ...552L..35Z} Zhang, B., \& M{\'e}sz{\'a}ros, P.\ 2001, \apjl, 552, L35
\bibitem[Zhang et al.(2007)]{2007ApJ...655L..25Z} Zhang, B., Zhang, B.-B., Liang, E.-W., et al.\ 2007, \apjl, 655, L25
\bibitem[Zhang et al.(2009)]{2009ApJ...703.1696Z} Zhang, B., Zhang, B.-B., Virgili, F.~J., et al.\ 2009, \apj, 703, 1696
\bibitem[Zhang et al.(2014)]{2014ApJ...787...66Z} Zhang, B.-B., Zhang, B., Murase, K., Connaughton, V., \& Briggs, M.~S.\ 2014, \apj, 787, 66
\end{thebibliography}
\end{document}